\documentclass[aps,prl,preprint,groupedaddress,showkeys]{revtex4-2}

\usepackage{cancel}
\usepackage{amsmath}
\usepackage{amsfonts}
\usepackage{graphicx}
\newcommand{\f}[1]{\boldsymbol{#1}}
\newcommand{\C}[1]{\mathcal{#1}}
\newcommand{\ti}[1]{\tilde{#1}}
\DeclareMathOperator{\ext}{\mathbf{d}}

\DeclareMathOperator{\mdot}{\qquad .}
\DeclareMathOperator{\mcomma}{\qquad ,}
\newcommand{\eqdef}{\;:=\;}
\newcommand{\eqimp}{\;\stackrel{!}{=}\;}

\begin{document}


\title{Square-torsion gravity, dark matter halos and the baryonic Tully-Fisher relation}


\author{Elias A. S. M\'egier}
\email[E-mail address: ]{elias.megier@unimi.it}
\altaffiliation{INFN Sez. di Milano, Via Celoria 16, 10133 Milano, Italy}
\affiliation{Dipartimento di Matematica F. Enriques, Universit\`a degli Studi di Milano, Via C. Saldini 50, 20133 Milano, Italy}
\thanks{My gratitude goes to Sergio L. Cacciatori and Livio Pizzocchero for their steadfast trust and to Francesco Cremona for our inspiring conversations.}

\date{\today}

\begin{abstract}
\emph{Square-torsion gravity} is applied to the long standing \emph{dark matter problem}. In this context the theory reduces to General Relativity complemented by a \emph{dark stress-energy tensor} due to the \emph{torsion} of spacetime and is studied under the simplifying assumption of \emph{spherical symmetry}. The dark stress-energy tensor is found to satisfy an anisotropic structure equation. In vacuum this is shown to be equivalent to a wave equation with sources. A natural class of exact solutions is found which explicitly perturbs any seed spacetime metric by a conformal factor satisfying a (1+1)-dimensional wave equation. This leads to the concept of \emph{dark coating}. The static solutions are then used to construct structures that model \emph{dark matter halos} surrounding baryonic bodies.  In the Newtonian r\'egime the baryonic mass $m_b$ and the flat rotation curve velocity $v_f$ are found to be \emph{related} by the \emph{baryonic Tully-Fisher relation} $m_b\propto v_f^4$. The present work proposes thus a possible theoretical motivation of this hitherto purely empirical result.  The example of a dark halo on the Schwarzschild geometry is made as a toy model for a galaxy. All qualitative an quantitative features of galactic rotation curves are recovered. A dark halo surrounding a Schwarzschild black hole is found to possess a boundary of staticity called \emph{torsion sphere} placed between the photon sphere and the event horizon. The phenomenon of \emph{dark radiation} is briefly exposed. The way for \emph{cosmological} applications is then opened by showing how Hubble expansion is a natural feature of the theory.
\end{abstract}


\maketitle

\section{\label{Intro}Introduction}
The present work investigates an alternative theory of gravity closely related to General Relativity. Both theories have a common mathematical and aesthetic foundation: gravitational phenomena are described by the \emph{geometry} of spacetime. The main difference between these sibling theories is that while Einstein's theory derives from an action functional containing only the \emph{Riemann curvature tensor}, Square-torsion Gravity completes such functional by a term quadratic in the \emph{torsion tensor}. Our main motivation for the investigation of alternative theories of gravitation is the hope of incorporating into the same framework phenomena that have a long standing problematic interpretations in terms of General Relativity. This article focusses on the so called \emph{dark matter problem}.

\subsection{\label{DProb}The dark matter problem}
The phenomenon of dark mater has been know for almost a century now, its scientific history is thus a subtle and fascinating matter in itself. We hereby give a very concise historical perspective and invite the reader to refer to the very complete historical review \cite{DarkHist} and its rich references. 

\subsubsection{Early history}
The first signs of a new open problem arose as early as 1933, when Zwicky estimated the mass of the Coma galaxy cluster using Virial's theorem and the relation between redshift and distance. He obtained a very high mass-to-light ratio. This suggested the presence of dark matter in some form. This fact could as well be explained by the possible  non equilibrium of the cluster. For this reason during the 1950s much effort was focussed on determining mass-to-light ratios of various clusters. It became however more and more difficult to reject the presence of dark matter using the hypothesis of non equilibrium. The main perplexity of the researchers came from the great age of the galaxies forming the clusters. The idea thus formed that 90 to 99\% of the mass of these clusters could be in the form of extragalactic material. Indeed by the 1970s the relaxation process of clusters seemed to support this idea. Further enquiries of x-ray emissions put however very stringent upper bounds to intergalactic hot gas, orders of magnitude less than the gas required to explain gravitational observations. This opened the way to more exotic possibilities.

Then another major observational difficulty arose, namely that of galactic rotation curves. The velocity of the stars orbiting around the galactic centre as a function of the distance from the centre itself can in principle be used to infer the mass distribution in the galaxy. The problem was that, assuming the mass distribution closely follows the luminosity distribution, one expects rotation curves that decrease in a Keplerian way after some critical value of the distance from the galactic centre. What one observes is typically very different, namely \emph{flat rotation curves}. By the mid 1970s enough evidence were collected from radio astronomy to convince many that the outskirts of galaxies should contain very large amounts of invisible mass.

Naturally, many hypotheses began to be formulated regarding the exact nature of this dark matter. Some suggested it could comprise planets, brown dwarfs, red dwarfs, neutron stars, and black holes. Another paradigm was that of weakly interacting fundamental particles yet to be identified. However, gravitational micro-lensing observations of galactic halos seemed to rule out the first possibility. 

The term \emph{dark matter} has nowadays become synonymous of particles that may account for the missing matter in our Universe.

Taking this point of view, one has then two possibilities to model such particles, as a \emph{hot} or as a \emph{cold} gas. These two models lead to very different structure formation on large scales. The most widely accepted model is that of cold dark matter, ruling therefore out candidate particles of the standard model such as relativistic neutrinos. This idea gained consensus as early as the end of the 1980s.

The picture handed to the 21${}^{st}$ century is hence that of a pervasive dark dust, of yet unknown nature. In spite of all the refined techniques which have been devised during the last decades and in spite of the dedicated search undertaken by many prized experimentalists of these new particles, no successful candidate was found.

\subsubsection{Recent developments}

The last two decades shed a new light onto two additional elements worthy of meditation:
\begin{itemize}
	\item The baryonic mass $m_b$ of a galaxy is strongly correlated to the magnitude of the flat rotation velocity $v_f$. This relation, known as \emph{baryonic Tully-Fisher relation}, takes the simple form $m_b\propto v_f^4$. As reported in \cite{BarTully} there does not seem to be any dependence on other properties like galactic size or surface brightness. This points out a possible deeper connection between the baryons and the physics which determines $v_f$.
	
	\item Data from hundreds of very diverse spiral galaxies seem to point out a fundamental correlation between the baryonic matter distribution and that of dark matter, irrespective of the fact that the galaxy be baryon or dark matter dominated \cite{AccelRot}.
	
\end{itemize}  A theoretically oriented mind could hence be tempted to embrace the possibility that the phenomena hitherto analysed using the dark matter paradigm might be explained by \emph{fundamental laws of gravitation yet to be understood}. We hereby explore this very possibility and attempt a purely \emph{geometrical} description rather than one derived from fundamental particles. It is however important to note that there are fundamental observational differences between galaxies and clusters of galaxies. To name one, the baryonic Tully-Fisher relation can only be observed in galaxies and not in galactic clusters. In spite of this fact, recent observational research indicates that there might be a nearly fundamental constant which parametrises the dark-baryonic matter interplay. We invite the interested reader to refer to \cite{NatureD} and its rich references. 

\subsection{\label{Square-torsion gravity}Square-torsion gravity?}
All this being said, a specific theory has to be chosen. A promising candidate is the so called \emph{square-torsion theory of gravity}. It was recently shown in \cite{Arawi} that in this theory some components of the \emph{torsion tensor} remain to be fixed a priori as though they were \emph{external sources}. This analogy with matter becomes mathematically \emph{exact} for a spacetime with vanishing spin density. The free degrees of freedom then appear in the Einstein equations as a \emph{stress-energy tensor due to torsion}, hereafter christened ``dark stress-energy tensor". Given that such stress-energy can take very diverse forms, e.g. that of \emph{dust}, this theory is a natural choice when trying to geometrise dark matter.

\section{Square-torsion Gravity in brief}
We hereby give a very brief overview of the main features of Square-torsion Gravity, we refer again to \cite{Arawi} and references therein for a more detailed account. 
\subsection{Main ingredients and notation}
Our mathematical model for spacetime is that of an orientable four-dimensional differentiable manifold $\C{M}$ endowed with a Lorentzian metric $\f{g}$. We use a local orthonormal basis $\f{e}:=\{\f{e}_{I}\}_{I\in\{0,1,2,3\}}$ of the tangent space $\C{TM}$. Let us call $\f{\theta}$ a local basis of the cotangent space $T^*\C{M}$ defined by $\f{\theta}^{I}\left(\f{e}_J\right)=\delta^I_J$, or in \emph{abstract index notation} $\theta^I{}_\mu e_J{}^\mu=\delta^I_J$,
where Greek indices run from $0$ to $3$ and denote ordinary indices of $T\C{M}$ and $T^*\C{M}$ respectively. With this structure, one point-wise recovers all the machinery of special relativity. In particular, each affine space is endowed with a Minkowski metric $g_{IJ}= \eta_{IJ} = diag(-1; +1; +1; +1)$  constant
throughout the computations. The inverse frame field $\theta^I{}_\mu$ pulls back this metric metric $g_{IJ}$ to define a metric on
spacetime. The inertial structure of spacetime is encoded in an
\emph{affine connection}, equivalent to the definition of a \emph{covariant derivative} of our reference frame as
\begin{equation}
\nabla_\mu \f{e}_I{}\eqdef \f{e}_J{}\:\omega^J{}_{I\mu}\mdot
\end{equation}
The \emph{torsion} $2$-form is defined as
\begin{equation}
T^I{}_{\mu\nu}:=2\theta^I{}_{[\nu,\mu]}+2\omega^I{}_{S[\mu}\theta^S{}_{\nu]}\label{deftor}\mcomma
\end{equation}
while the \emph{Riemann curvature} 2-form is defined as
\begin{equation}
R^{IJ}{}_{\mu\nu} \eqdef\left[\nabla_\mu,\nabla_\nu\right]^I{}_J= 2\omega^{IJ}{}_{[\nu,\mu]}+2\omega^{I}{}_{S[\mu}\omega^{SJ}{}_{\nu]}\mcomma
\end{equation}where we use the common bracket notation to denote a skew-symmetrisation of indices.
\paragraph{Relation between the Levi-Civita and a general torsionful connection:} It is well known that there is a unique metric-compatible connection which is torsion-free. This is called Levi-Civita connection and we denote it by $\Gamma^I{}_{J\mu}$. From definition \eqref{deftor} of the torsion $2$-form one can easily demonstrate that
\begin{equation}
\Gamma_{IJK}=\frac{1}{2}[\f{e}_K,\f{e}_J]_I+\frac{1}{2}[\f{e}_I,\f{e}_K]_J-\frac{1}{2}[\f{e}_J,\f{e}_I]_K\mdot\label{LC}
\end{equation} Notice how we lowered indices for practical purposes using the metric $g_{IJ}$ and that the brackets are \emph{Lie brackets}. In the same way
for a generic torsionful connection $\omega^I{}_{J\mu}$ we can relate it to $\Gamma^I{}_{J\mu}$ as follows
\begin{equation}
\omega_{IJK}=\Gamma_{IJK}+\Delta_{IJK}\label{decomp}
\end{equation}
where
\begin{equation}
\Delta_{IJK}=\frac{1}{2}\left(T_{KIJ}+T_{IKJ}-T_{JKI}\right)
\end{equation} is called the \emph{contorsion tensor}. 
\subsection{Action functional and equations of motion}
The action we consider is 
\begin{equation}
S=\frac{1}{16\pi G}
\int\limits_{\mathcal{M}}
d^4x\: det(\theta)\left(\frac{1}{2}T^I{}_{JK}T_I{}^{JK}-R^{IJ}{}_{IJ}\right)+\int\limits_{\mathcal{M}}
d^4x\:\C{L}_s\label{action}
\end{equation}	where $\C{L}_s$ is the Lagrangian density for all external sources coupled to geometry and $d^4x\: det(\theta)$ the volume form. 
This is the Einstein-Hilbert action complemented by a quadratic dynamical term for the torsion of spacetime. It is conceived as a \emph{first order formalism} action principle, in which the connection $\omega^{IJ}{}_\mu$ and the  tetrad $\theta^I{}_\mu$ are independent variables. The equations of motion derived from this action by its variation with respect to the connection yield equations that have \emph{intrinsic angular momentum} as source. In most macroscopic applications, when dealing with perfect fluids or more general fluids that do not include covariant derivatives in their structure equation, we can assume said spin density to vanish. This in turn means that 
\begin{align}
T^S_{\phantom{S}IS}&=0\nonumber\\
T_{[IJK]}&=0\mdot
\end{align}The only non vanishing irreducible component of the torsion tensor is thus that which is traceless and has vanishing antisymmetric part. This irreducible component shall be denoted by $\C{T}$.

Let us now come to the equations of motion having the \emph{stress-energy tensor} of matter $P^{IJ}$ as source. As said, we are interested in non-nematic phases of macroscopic matter and cosmological fluids. In this setting the equations of motion derived from the action functional by its variation with respect to the tetrad are simply
\begin{align}
\frac{1}{2}\C{T}^I{}_{ST}\C{T}^{JST}+\C{G}^{IJ}=8\pi\:G\:P^{IJ}\label{Main3}\mcomma
\end{align} where $\C{G}^{IJ}$ is the usual Einstein tensor. This is very interesting because we still have 16 free torsion components (up to gauge choice) and, if one interprets these equations as the Einstein equations, the tensor
\begin{equation}
D^{IJ}\eqdef-\frac{1}{16\pi G}\C{T}^I{}_{ST}\C{T}^{JST}
\end{equation}
plays the role of a \emph{dark content}. Indeed, taking the covariant Levi-Civita divergence of \eqref{Main3} we find
\begin{align}
P^{IS}{}_{;S}+D^{IS}{}_{;S}=0\mdot
\end{align}
If we reason in terms of the Levi-Civita connection we are hence forced to consider $D^{IJ}$ as a stress-energy tensor due to the torsional degrees of freedom of our geometry because it is included in the conservation of energy and momentum.
Notice that isolated point particles follow geodesics, because semi-classical matter fields are governed by their own equations of motion, which in the eikonal approximation imply geodesic motion.

It is useful to work with a local decomposition of the torsion tensor in terms of its ``electric'' and ``magnetic'' parts. To this end we choose a preferred time direction $\theta^0{}_\mu$ and denote spacelike indices with lowercase Latin indices running from $1$ to $3$. We can thus uniquely define
\begin{align}
T^I{}_{i0}&=E^I{}_i\nonumber\\
T^I{}_{ij}&=\epsilon_{ij}{}^\kappa B^I{}_\kappa\mdot
\end{align}In turn, it is useful to decompose the $3$-dimensional matrices $E_{ij}$ and $B_{ij}$ into their \emph{traceless} symmetric and skew-symmetric parts:
\begin{align}
e_{ij}&=E_{(ij)}\nonumber\\
b_{ij}&=B_{(ij)}\nonumber\\
\epsilon_{ij}{}^k\varepsilon_\kappa&=E_{[ij]}\nonumber\\
\epsilon_{ij}{}^k\beta_\kappa&=B_{[ij]}\mdot
\end{align}Using the equations of motion one can finally write:
\begin{align}
	2\pi G D_{00}&=-\varepsilon_i\varepsilon^i+\beta_i\beta^i\\
	4\pi G D_{0i}&=-e_i{}^j\beta_j-b_i{}^j\varepsilon_j \label{misprint}\\
	8\pi G D_{ij}&=E_i{}_sE_j{}^s-B_i{}_sB_j{}^s\mdot
	\end{align}(Equation \eqref{misprint} corrects a misprint in \cite{Arawi}.)
\section{The assumption of spherical symmetry}
From what we have seen in the previous section, the structure of $\C{D}$ remains possibly quite complex. For this reason we specialise our discussion to configurations with \emph{spherical symmetry}, which, in addition to being a strongly simplifying assumption, is of great physical interest. 

What do we mean by spherical symmetry in this context? We adopt the common definition according to which a Lorentzian manifold $(\C{M},\f{g})$ in \emph{spherically symmetric} if it has the group of three dimensional rotations $SO(3)$ as \emph{isometry group}. We moreover require that $SO(3)$ should act in such a way that its orbits be \emph{two dimensional spacelike surfaces}.

It can be demonstrated, see e.g. \cite{Strau}, that the requirement of spherical symmetry formulated in this fashion implies that $(\C{M},\f{g})$ be locally a \emph{warped product} $\C{M}=\tilde{\C{M}}\times_R S^2$, where $(\tilde{\C{M}},\tilde{\f{g}})$ is a two dimensional Lorentzian manifold and $(S^2,{\f{g}_{S^2}})$ denotes the two-sphere. $R:\tilde{\C{M}}\rightarrow\mathbb{R}$ is the so called \emph{warping function}, which can be though as ``radius function" given that the warped product metric by definition takes the form
\begin{equation}
	\f{g}=\tilde{\f{g}}+R^2\:\f{g}_{S^2}\mdot
\end{equation}
Let us now turn to the torsion tensor. Spherical symmetry implies the impossibility to single out a preferred element belonging to the distribution of vectors tangent to $S^2$. If e.g. we introduce spherical angular coordinates so that $\f{g}_{S^2}=\ext\vartheta\otimes\ext\vartheta+\sin^2\vartheta\ext\varphi\otimes\ext\varphi$ and use an adapted orthonormal frame in which $\f{e}_2\eqdef R^{-1}\partial_\vartheta$ and $\f{e}_3\eqdef R^{-1}\sin\vartheta^{-1}\partial_\varphi$ we must have
\begin{align}
	(\varepsilon^i)_{i=1,2,3}&=:\:(\varepsilon,0,0)\nonumber\\
	(\beta^i)_{i=1,2,3}&=:\:(\beta,0,0)\nonumber\\
	(e^{ij})_{i,j=1,2,3}&=:\:\text{diag}(e,-e/2,-e/2)\nonumber\\
	(b^{ij})_{i,j=1,2,3}&=:\:\text{diag}(b,-b/2,-b/2)
\end{align}(recall that $e^{ij}$ and $b^{ij}$ are traceless).
The dark stress energy thus becomes
\begin{align}
2\pi\:G D_{00}&=-\varepsilon^2+\beta^2\nonumber\\
(4\pi\:G D_{0i})_{i=1,2,3}&=-(e\beta+b\:\varepsilon,0,0)\nonumber\\
(8\pi\:G D_{ij})_{i,j=1,2,3}&=\text{diag}(e^2-b^2,\frac{e^2-b^2}{4}+\varepsilon^2-\beta^2,\frac{e^2-b^2}{4}+\varepsilon^2-\beta^2)\label{details}\mdot
\end{align}Notice how both the energy density component and pressure can have a priori any sign. Let us then define $\rho\eqdef D_{00}$, $P\eqdef D_{11}$ and $P_\bot\eqdef D_{22}$. The dark stress-energy tensor then satisfies the structure equation
\begin{equation}
	P-\rho=4 P_\bot\mdot\label{stateeq}
\end{equation}

\subsection{\label{statrem}Remark on staticity}
In view of the following applications it is important to dwell a moment on the expression we obtained for the energy current density component $J\eqdef D_{01}$. At first sight it might seem completely innocuous. It has however important implications e.g. for the study of \emph{vacuum configurations}. In vacuum the stress-energy tensor due to matter vanishes. It then becomes the only source for the Einstein tensor. This implies that the static character of vacuum solutions is governed by $J$. We hence state the following\\

\paragraph{\label{Jcond}Compatibility condition on $J$}
\begin{itemize}
	\item if a dark stress-energy tensor has $J=0$ then $\rho$ and $P$ have \emph{opposite sign} (or vanish)
\end{itemize}
	or equivalently
\begin{itemize}
	\item if a dark stress-energy tensor has $\rho$ and $P$ with the same sign (and non vanishing) then $J\neq0$.
\end{itemize}

The two statements being logically equivalent, let us demonstrate the second point. We start by assuming that $\rho>0$ and $P>0$. Expression \eqref{details} then implies $|\beta|>|\varepsilon|$ and $|b|>|e|$. We can thus write the little chain of inequalities $|e\beta|=|e||\beta|>|e||\varepsilon|>|b||\varepsilon|=|b\varepsilon|$. From this we deduce that $J\neq0$. The case where both $\rho$ and $P$ are negative can be demonstrated in the very same way by swapping the role of the electric and magnetic components.

In Square-torsion gravity one can hence have spherically symmetric vacuum solutions that are \emph{not} static. This fact should be clear if one considers that the present theory is equivalent to General Relativity in the presence of a fluid. Birkhoff's theorem does hence not apply in this setting.

\subsection{Remark on gauge freedom}
Due to the fact that $D$ is quadratic in the torsion tensor, the relation between the components of torsion and those of the dark stress-energy tensor is not invertible. The choice of torsion tensor remains thus not unique. This freedom can be parametrised e.g. by two hyperbolic angles $(\eta,\tau)$ as follows:
\begin{itemize}
	\item for $\rho>0$ and $P>0$
	\begin{align}
	\varepsilon&=\pm\sqrt{2\pi\:G\rho}\:\sinh\eta\nonumber\\
	\beta&=\pm\sqrt{2\pi\:G\rho}\:\cosh\eta\nonumber\\
	e&=\pm 2\sqrt{2\pi\:GP}\:\sinh\tau\nonumber\\
	b&=\pm 2\sqrt{2\pi\:GP}\:\cosh\tau\nonumber\\
	\end{align}
	\item the case of different signs is obvious.
\end{itemize}

These hyperbolic angles are however not independent. We still have to solve for the energy current density $J$. It is easy to see e.g. that if $J=0$ then $\tau=\pm\eta$.

\section{The equations of motion in vacuum}
In order to compute the equation of motion for our spacetime geometry in vacuum we now proceed with the computation of the Einstein tensor for our warped product. To this end, let us introduce a pair of coordinates $(t,x)$ on $\tilde{\C{M}}$ and an adapted local orthonormal frame $\f{\ti{\theta}}\eqdef(\f{{\theta}}^{\tilde{I}})_{\tilde{I}=0,1}$, so that a tilde singles out directions tangent to $\tilde{\C{M}}$. Let then a hat single out directions tangent to $S^2$ in an analogous way $\f{\hat{\theta}}\eqdef(\f{{\theta}}^{\hat{I}})_{\hat{I}=2,3}$. Let us also change variables from the warping function $R$ to its logarithm $\psi\eqdef\log(R)$.
It is a standard task to compute the Riemann and Ricci tensors of the Levi-Civita connection associated to such warped product, we refer again to \cite{Strau} for more details. Let $\tilde{R}^{\ti{I}}{}_{\ti{J}}$ be the components of the Ricci tensor of $\f{\ti{g}}$, with this notation one finds the following expression for the components $R^I{}_J$ of the Ricci tensor of $\f{g}$:
\begin{align}
	R^{\ti{I}}{}_{\ti{J}}&=\tilde{R}{}^{\ti{I}}{}_{\ti{J}}-2\:\psi_{;}{}^{\ti{I}}{}_{\ti{J}}-\:2\psi^{,\ti{I}}\psi_{,\ti{J}}\nonumber\\
	R^{\ti{I}}{}_{\hat{J}}&=0\nonumber\\
	R^{\hat{I}}{}_{\hat{J}}&=\left(\mathrm{e}^{-2\psi}-\:\psi_{;}{}^{\ti{S}}{}_{\ti{S}}-2\:\psi^{,\ti{S}}\psi_{,\ti{S}}\right)\delta^{\hat{I}}_{\hat{J}}\mcomma
	\end{align}where we used the fact that the Ricci tensor of $S^2$ takes the simple form $\frac{1}{2}\delta^{\hat{I}}{}_{\hat{J}}$.

Let us now translate the structure equation \eqref{stateeq} to the components of the Ricci tensor. In terms of the Einstein tensor we clearly have
\begin{equation}
	\C{G}^{\ti{S}}{}_{\ti{S}}=2\:\C{G}^{\hat{T}}{}_{\hat{T}}
\end{equation}
and thus
\begin{equation}
	2\:{R}^{\ti{S}}{}_{\ti{S}}=\:{R}^{\hat{T}}{}_{\hat{T}}\mdot
\end{equation}This is just the right balance to eliminate the term quadratic in $\psi_{,\ti{S}}$. We find
\begin{equation}
	\psi^{\ti{S}}{}_{\ti{S}}=\ti{R}^{\ti{S}}{}_{\ti{S}}-\mathrm{e}^{-2\psi}\mdot\label{step}
\end{equation} The last expression can be formulated in an even tamer way by remembering that the Lorentzian metric $\f{\ti{g}}$ is two-dimensional and is thus locally equivalent to a conformally flat metric $\mathrm{e}^{2\phi}\left(-\ext t\otimes\ext t + \ext x\otimes\ext x\right)$. The curvature scalar $\C{R}^{\hat{T}}{}_{\hat{T}}$ associated to such metric can thus be expressed as
\begin{equation}
	{R}^{\hat{T}}{}_{\hat{T}}=-2\:\phi_{;}^{\ti{S}}{}_{\ti{S}}\mdot
\end{equation}We can thus rephrase \eqref{step} as
\begin{equation}
	\left(-\partial^2_t+\partial^2_x\right)(\psi+2\phi)=-\mathrm{e}^{2\left(\phi-\psi\right)}
\end{equation}and changing variables to $\Sigma\eqdef2\phi+\psi$ and $\Delta\eqdef\phi-\psi$ we finally come to the result
\begin{equation}
	\Box\:\Sigma=-\mathrm{e}^{-2\Delta}\label{master}\mdot
\end{equation}Notice that this equation, together with the compatibility condition \ref{Jcond} on $J$, is the only we need to solve. This is due to the fact that, apart from the structure equation, $\rho$ and $P$ can take any desired form descending from a solution of \eqref{master}. This is exactly what we expected, in spherically symmetric Square-torsion Gravity one has one free parameter in the dark stress-energy tensor which can be used to model the observed dark matter density.

An important question then arises. Is there any such parametrisation of dark matter which can be theoretically preferred?

\section{Conformal coatings}
Our evolution equation \eqref{master} is a (1+1)-dimensional wave equation with source. The wave function $\Sigma$ is independent from the source function $\Delta$. This means that, given any seed solution $(\Sigma_0,\Delta_0)$ of \eqref{master} the couple $(\Sigma_0+3\:\varsigma,\Delta_0)$ is still a solution if 
\begin{equation}
	\Box\:\varsigma=0\mdot
\end{equation}
What does $\varsigma$ represent in physical terms? Recall that we had expressed $\f{g}$ as
\begin{equation}
	\f{g}=\mathrm{e}^{2\phi}\left(-\ext t\otimes\ext t+\ext x \otimes\ext x\right)+\mathrm{e}^{2\psi}\f{g}_{S^2}
\end{equation} which, rewritten in terms of our new variables $(\Sigma,\Delta)$ reads
\begin{equation}
\f{g}=\mathrm{e}^{\frac{2}{3}\Sigma}\left(\mathrm{e}^{\frac{2}{3}\Delta}\left(-\ext t\otimes\ext t+\ext x \otimes\ext x\right)+\mathrm{e}^{-\frac{4}{3}\Delta}\f{g}_{S^2}\right)\mdot
\end{equation} It is now clear that the above mentioned symmetry in the space of solutions has a direct geometrical interpretations as \emph{harmonic conformal rescalings} of solutions. This leads naturally to the concept of \emph{conformal coating}.

\subsection{Conformal coatings in vacuum}
If we have a seed spherically symmetric solution $\f{g}_0$ satisfying the vacuum equation \eqref{master} any metric
\begin{align}
	&\f{g}=\mathrm{e}^{2\:\varsigma}\f{g}_0\nonumber\\
	&\Box_{\tilde{g}_{{}_0}}\varsigma=0\label{wave}
\end{align}is also a solution to \eqref{master}, provided that the compatibility condition on $J$ be verified.

In \eqref{wave} we have emphasised the fact that $\varsigma$ satisfies the wave equation for the \emph{two-dimensional Lorentzian metric $\f{\ti{g}_{{}_0}}$}. In particular, it is immediate to write local expressions of solutions to \eqref{wave} by means of D'Alembert's formula
\begin{equation}
	2\:\varsigma(t,x)=\varsigma(0,x+t)+\varsigma(0,x-t)+\int\limits^{x+t}_{x-t}ds\:\partial_t\varsigma(0,s)\mcomma
\end{equation}which shows that sufficiently regular and bounded initial perturbations $\varsigma(0,x)$ remain such in any chart. This class of \emph{exact metric perturbations} is hence of great physical interest. This is arguably the most natural class of solutions to \eqref{master} because it descends from a natural \emph{symmetry} of said equation. If we have a physically preferred solution $\f{g}$, equation \eqref{wave} provides its ``dark coating", so to speak. Moreover, we are about to see that this coating method extends directly to \emph{non empty Einsteinian solutions}.

\paragraph{Example regarding more general vacuum scenarios:}Let us make an explicit example in order to show how dangerous more general perturbations of \eqref{master} can be. Let us choose $\f{g}_0$ very simply to be the Minkowsky metric. If we perturb also $\Delta_0$ with a perturbing function $\delta$ our equation becomes
\begin{equation}
	3\;\Box\:\varsigma=-\frac{1}{x^2}\left(\mathrm{e}^{2\delta}-1\right)\mdot
\end{equation}Let us now assume $2\:d\eqdef\left(\mathrm{e}^{2\delta}-1\right)$ to depend linearly on $\varsigma$ so as to have
\begin{equation}
	-\partial^2_t\varsigma=-\partial^2_x\varsigma+\frac{\alpha}{x^2}\varsigma=:\:\C{L}_\alpha\varsigma\label{Schr}
\end{equation}for some $\alpha\in\mathbb{R}$. $\C{L}_\alpha$ is evidently a Schr\"odinger  operator for the potential $\alpha/x^2$, generally known as \emph{Calogero potential}. It suffices then to study the time independent Schr\"odinger equation associated to $\C{L}_\alpha$. Loosely speaking, from \eqref{Schr} one deduces that a negative part of the spectrum of $\C{L}_\alpha$ will result in unstable solutions. This problem is studied in great detail in \cite{Calogero}. In particular, for $\alpha<-1/4$ such negative spectrum does not only exist, but is unbounded from below! Notice that we made an \emph{exact} consideration, no linearisation was needed, we just chose a simple \emph{ad hoc} perturbing direction in an appropriately well behaved subspace of solutions to \eqref{master}.  

\subsection{General conformal coatings}
For ease of discussion and future use, let us preliminarily express the Ricci and Einstein tensors for a conformally coated metric in function of their uncoated relatives. To this end it suffices to express a coated orthonormal frame of our choice in therms of its uncoated relative as $\f{\vartheta}^{\ddot{I}}=\mathrm{e}^{\varsigma}\f{\theta}^{I}$, thus denoting coated indices with a diaeresis. Using the vanishing torsion condition one can write
\begin{equation}
	\ext\f{\vartheta}^{\ddot{I}}=\ext\varsigma\wedge\f{\theta}^I-\mathrm{e}^{\varsigma}\:\f{\Gamma}^I{}_J\wedge\f{\theta}^J
\end{equation}from which we deduce that
\begin{equation}
	\f{\Gamma}^{\ddot{I}}{}_{\ddot{J}}=\varsigma_{,J}\f{\theta}^I-\varsigma^{,I}\f{\theta}_J+\f{\Gamma}^I{}_J\label{conn}
\end{equation}and this enables us to express the Riemann curvature two-form as
\begin{align}
	\f{R}^{\ddot{I}}{}_{\ddot{J}}&=\ext\f{\Gamma}^{\ddot{I}}{}_{\ddot{J}}+\f{\Gamma}^{\ddot{I}}{}_{\ddot{S}}\wedge\f{\Gamma}^{\ddot{S}}{}_{\ddot{J}}\nonumber\\
	&=\f{R}^I{}_J+\varsigma_{;JS}\:\f{\theta}^S\wedge\f{\theta}^I-\varsigma_{;}{}^I{}_{S}\:\f{\theta}^S\wedge\f{\theta}_J+\nonumber\\
	&\phantom{=}+\varsigma_{,J}\f{\theta}^I\wedge\ext\varsigma-\varsigma_{,S}\:\varsigma^{,S}\f{\theta}^I\wedge\f{\theta}_J-\varsigma^{,I}\f{\theta}_J\wedge\ext\varsigma\mdot
\end{align} From this expression we can read off the components of the Riemann tensor and arrive at the Ricci
\begin{align}
	\mathrm{e}^{2\;\varsigma}R^{\ddot{I}}{}_{\ddot{J}}&=R^I{}_J-2\:\varsigma_{;}{}^{I}{}_{J}-\varsigma_{;}{}^{S}{}_{S}\:\delta^I_J+2\;\varsigma^{,I}\varsigma_{,J}-2\;\varsigma^{,S}\varsigma_{,S}\:\delta^I_J\nonumber\\
	&=:R^I{}_J\:+\stackrel{\varsigma}{R}{}^I{}_J\mdot\label{coatRi}
\end{align}This expression for the Ricci tensor is of great relevance because it shows how an extended version of our result on conformal coatings holds:

If we have a spherically symmetric seed square-torsion gravitational solution $\f{g}_0$, possibly describing an Einsteinian spacetime filled with matter, then
\begin{align}
&\f{g}=\mathrm{e}^{2\varsigma}\f{g}_0\nonumber\\
&\Box_{\tilde{g}_{{}_0}}\varsigma=0\label{genwave}
\end{align}is also a square-torsion gravitational solution, provided that the compatibility condition on $J$ be verified.

Let us demonstrate this fact by checking whether the structure equation holds for $\stackrel{\varsigma}{R}{}^I{}_J$. Notice first of all that because of spherical symmetry $\varsigma:\ti{\C{M}}\rightarrow\mathbb{R}$ and one clearly has $\varsigma_{,\hat{I}}=0$. This in turn implies that $\Box_{\ti{g}{}_{{}_0}}\varsigma=0$ is equivalent to $\varsigma_{;}{}^{\ti{S}}{}_{\ti{S}}=0$. We hence have $\varsigma_{;}{}^{{T}}{}_{{T}}=\varsigma_{;}{}^{\hat{S}}{}_{\hat{S}}$, while $\varsigma^{,T}\varsigma_{,T}=\varsigma^{,\ti{S}}\varsigma_{,\ti{S}}$. We can thus write the following chain of equalities:
\begin{align}
	2\;\stackrel{\varsigma}{R}{}^{\ti{S}}{}_{\ti{S}}&=-4\:\varsigma_{;}{}^{\ti{S}}{}_{\ti{S}}-4\:\varsigma_{;}{}^{T}{}_{T}+4\;\varsigma^{,\ti{S}}\varsigma_{,\ti{S}}-8\:\varsigma^{,T}\varsigma_{,T}\nonumber\\
	&=-4\:\varsigma_{;}{}^{\hat{T}}{}_{\hat{T}}-4\;\varsigma^{,\ti{S}}\varsigma_{,\ti{S}}\nonumber\\
	&=-2\:\varsigma_{;}{}^{\hat{T}}{}_{\hat{T}}-2\:\varsigma_{;}{}^{S}{}_{S}-4\:\varsigma^{,S}\varsigma_{,S}\nonumber\\
	&=\:\stackrel{\varsigma}{R}{}^{\hat{T}}{}_{\hat{T}}\mdot
\end{align}

From this expression of the Ricci tensor it is now easy to derive that of the Einstein tensor:
\begin{align}
\mathrm{e}^{2\varsigma}\C{G}^{\ddot{I}}{}_{\ddot{J}}&=\C{G}^I{}_J-2\:\varsigma_{;}{}^{I}{}_{J}+2\;\varsigma_{;}{}^{S}{}_{S}\:\delta^I_J+2\;\varsigma^{,I}\varsigma_{,J}+\varsigma^{,S}\varsigma_{,S}\:\delta^I_J\nonumber\\
&=:\C{G}^I{}_J\:+\stackrel{\varsigma}{\C{G}}{}^I{}_J\mdot\label{coatEin}
\end{align}

\subsection{Baryon conservation} 
From what we have hitherto shown, in the case of general conformal coatings the distinction between the dark components of the Einstein tensor and the rest is clear cut. Even more so thanks to the fact that the \emph{contracted Bianchi identity holds separately} for the two components. Let us express this fact more eloquently. Let $P^{IJ}$ be the total stress-energy tensor. As done before, let us write it as a sum of a \emph{baryonic} contribution $P_b{}^{IJ}$ and a dark one $D^{IJ}$. We then write our field equations as
\begin{equation}
	\mathrm{e}^{2\varsigma}\C{G}^{\ddot{I}\ddot{J}}=\C{G}^{IJ}\:+\stackrel{\varsigma}{\C{G}}{}^{IJ}=8\pi\;G\left(P_b^{IJ}+D^{IJ}\right)
\end{equation} and write the two separate equations
\begin{align}
	\C{G}^{IJ}&=8\pi\;GP_b^{IJ}\nonumber\\
	\stackrel{\varsigma}{\C{G}}{}^{IJ}&=8\pi\;GD^{IJ}\mdot
\end{align}This last step is meaningful because of the restricted identity
\begin{align}
	\stackrel{\varsigma}{\C{G}}{}^{\ddot{I}\ddot{K}}{}_{;\ddot{K}}\equiv0\mdot\label{specialBianchi}
\end{align}To prove this, let us first express the coated covariant divergence of a generic symmetric tensor $P^{\ddot{I}\ddot{J}}$ in terms of uncoated objects. By means of \eqref{conn} it is easy to see that
\begin{equation}
	P^{\ddot{I}\ddot{K}}{}_{;\ddot{K}}=\mathrm{e}^{-3\varsigma}\left(P^{IK}{}_{;K}+2\;P^{IK}\varsigma_{,K}-P^K{}_K\varsigma^{,I}\right)\mdot\label{coatdiv}
\end{equation}
 If we now insert the definition \eqref{coatEin} of $\stackrel{\varsigma}{\C{G}}{}^{\ddot{I}\ddot{J}}$ in terms of $\varsigma$ into \eqref{coatdiv} we have
 \begin{align}
 	\mathrm{e}^{3\varsigma}\stackrel{\varsigma}{\C{G}}{}^{\ddot{I}\ddot{K}}{}_{;\ddot{K}}&=-2\;\varsigma_{;}{}^{IK}{}_K+2\;\varsigma_{;}{}^S{}_S{}^I+2\left(\varsigma^{,I}\varsigma^{,K}\right)_{;K}+\left(\varsigma^{,S}\varsigma_{,S}\right)_{;}{}^I+\nonumber\\
 	&\phantom{=}-4\;\varsigma_;{}^{IK}\varsigma_{,K}+4\;\varsigma_;{}^S{}_S\varsigma^{,I}+4\;\varsigma^{,I}\varsigma^{,S}\varsigma_{,S}+2\;\varsigma^{,S}\varsigma_{,S}\varsigma^{,I}-6\;\varsigma^{;}{}^S{}_S\;\varsigma^{,I}-6\;\varsigma^{,S}\varsigma_{,S}\;\varsigma^{,I}\nonumber\\
 	&= -2\;\varsigma_{;}{}^{IK}{}_K+2\;\varsigma_{;}{}^S{}_S{}^I-2\;\varsigma_{;}{}^{IK}\;\varsigma_{,K}+2\;\varsigma_{;}{}^{KI}\;\varsigma_{,K}\nonumber\\
 	&=-2\;\mathrm{e}^{-\varsigma}\left(\mathrm{e}^{\varsigma}\left(\varsigma_{;}{}^{IK}-\varsigma_{;}{}^{KI}\right)\right)_{;K}\mdot
 	\end{align}
 	Notice now that we are using the uncoated Levi-Civita connection, which we explicitly expressed in \eqref{LC} as Lie brackets of frame fields. We use this explicitly and write
 	\begin{align}
 		\varsigma_{;IK}-\varsigma_{;KI}&=\varsigma_{,I;K}-\varsigma_{,K;I}\nonumber\\
 		&=\varsigma_{,IK}-\varsigma_{,KI}+\varsigma_{,S}\left(\Gamma_I{}^S{}_K-\Gamma_K{}^S{}_I\right)\nonumber\\
 		&=[\f{e}_K,\f{e}_I]_S\;\varsigma^{,S}+\varsigma_{,S}\left(\Gamma_I{}^S{}_K-\Gamma_K{}^S{}_I\right)\nonumber\\
 		&=\varsigma^{,S}\left([\f{e}_K,\f{e}_I]_S+\frac{1}{2}[\f{e}_K,\f{e}_S]_I+\frac{1}{2}[\f{e}_I,\f{e}_K]_S-\frac{1}{2}[\f{e}_S,\f{e}_I]_K+\right.\nonumber\\
 		&\phantom{=}\left.-\frac{1}{2}[\f{e}_I,\f{e}_S]_K-\frac{1}{2}[\f{e}_K,\f{e}_I]_S+\frac{1}{2}[\f{e}_S,\f{e}_K]_I\right)\nonumber\\
 		&= 0\mcomma
 	\end{align} and \eqref{specialBianchi} is thus proved.
 	
 	Thanks to the coated contracted Bianchi identity
 	\begin{equation}
 			{\C{G}}{}^{\ddot{I}\ddot{K}}{}_{;\ddot{K}}\equiv0 	
 	\end{equation}
 	we can finally express the \emph{coated baryonic conservation equation}
 	\begin{equation}
 		P_b^{\ddot{I}\ddot{K}}{}_{;\ddot{K}}=0\mdot\label{barcons}
 	\end{equation}It is hence physically meaningful to separate the two Einstein tensor components as above. Notice how this is a \emph{coated} expression, it hence depends on the dark stress-energy tensor. This is what in the present model gives rise to dark matter-related effects such as the departure from Keplerianity of rotation curves.
\section{Applications}
Now that we know how to construct spherically symmetric dark coatings on any base solution of our choice, we turn to physically relevant examples. The choice of a base is quite solid, it is sufficient to take any relevant Einsteinian setting and use it as a plinth.

\subsection{Static setting}
We shall now focus on static solutions. We will hence use a chart of $\ti{\C{M}}$ with coordinates $(t,r)$ in which we can express staticity by the existence of a timelike killing filed $K=\partial_t$.
\subsubsection{Spherically symmetric cluster}
Due to its conceptual and practical importance, we now turn to a Spherically symmetric cluster surrounded by vacuum. Such an idealisation could conceivably be used as a toy model for both spherically symmetric galactic clusters and galaxies.
We then imagine to have a continuum model of \emph{dust}. Let $\rho_{{}_0}$ be the rest frame dust mass density and $u^I$ the four-velocity field components of our dust continuum. The general form of the stress-energy tensor associated to such fluid is well known to be
\begin{equation}
	P^I{}_J=\rho_{{}_0}u^Iu{}_J\mdot
\end{equation}
Let us furthermore assume that this mesoscopic model arises from a microscopic model of point-like particles bound in approximately circular orbits. This fact can be minimally implemented by considering $u^I$ to be a \emph{random variable} on each $p\in\C{M}$ with a spherically symmetric probability measure giving rise to the following moments
\begin{align}
	<u^I>&=\frac{1}{\sqrt{1-v^2}}\:\delta^I{}_0\nonumber\\
	<u^{\ti{I}}u_{\ti{J}}>&=\frac{-1}{1-v^2}\:\delta^{\ti{I}}_{0}\delta_{\ti{J}}^{0}\nonumber\\
	<u^{\hat{I}}u_{\hat{J}}>&=\frac{{v^2}/{2}}{1-v^2}\:\delta^{\hat{I}}_{\ti{J}}
\end{align}where $v$ is the angular velocity of the dust particles with respect to our reference frame. This expression simply states that at each point one is equally likely to find a dust particle swarming in any angular direction, that the second moment of the radial component of the velocity is negligible with respect to that of the angular component (approximately circular orbits) and that the mesoscopic magnitude $v$ of the orbiting velocity and the rest frame mass density $\rho_{{}_0}$ are no random variables, this to stick to the essential features only. Notice that the macroscopic stress energy tensor $<P^I{}_J>$ retains its normalisation $P^S{}_{S}=-\rho_{{}_0}$ as it should. If we introduce the same notation used for the dark stress-energy tensor, the macroscopic stress-energy tensor components  due to baryons are then
\begin{align}
	\rho_b&=\frac{\rho_{{}_0}}{1-v^2}\nonumber\\
	P_b&=0\nonumber\\
	P_{b_\perp}&=\frac{\rho_{{}_0}{v^2}/{2}}{1-v^2}\nonumber\\
	J_b&=0\mdot
\end{align}

Let us then introduce Schwarzschild-like coordinates $(t,r)$ and an uncoated orthonormal frame such that $\f{\theta}^0=\mathrm{e}^{\Phi}\ext t$ and $\f{\theta}^1=\mathrm{e}^{\Lambda}\ext r$. The expression of the connection and curvature for a fluid-filled spherically symmetric spacetime can be found in any General Relativity textbook, we refer e.g. to \cite{Bible} for a very complete account. Applied to our case, one finds in particular
\begin{align}
	\Phi(r)&=\int\limits_0^r dx \frac{Gm(x)}{x(x-2Gm(x))}\nonumber\\
	\Lambda(r)&=-\frac{1}{2}\log\left(1-\frac{2Gm(r)}{r}\right)\nonumber\\
	m(r)&=\int\limits_0^r dx\: 4\pi\rho_b(x)x^2\nonumber\\
	\Gamma^0{}_{10}&=\Phi_{,1}=\mathrm{e}^{-\Lambda}\frac{d}{dr}\Phi\nonumber\\
	&=\frac{Gm(r)}{r^{3/2}(r-2Gm(r))^{1/2}}\nonumber\\
	\Gamma^2{}_{12}&=\Gamma^3{}_{13}=\frac{(r-2Gm(r))^{1/2}}{r^{3/2}}\mdot
\end{align}
\paragraph{Uncoated case:}Let us preliminarily discuss the uncoated case. The only missing element in our discussion is that of solving the Einstein equations for $P_{b_\perp}$. To this end, let us write the uncoated conservation of stress-energy and focus on the only non trivial component
\begin{equation}
	P_b^{1S}{}_{;S}=0\mdot
\end{equation}More explicitly one has
\begin{align}
	P_b^{1S}{}_{;S}&=P_b^{10}{}_{;0}+P_b^{11}{}_{;1}+P_b^{12}{}_{;2}+P_b^{13}{}_{;3}\nonumber\\
	&=\left(P_b^{00}+P_b^{11}\right)\Gamma^0{}_{10}+P_b^{11}{}_{;1}+P_b^{11}\left(\Gamma^{2}{}_{12}+\Gamma^{3}{}_{13}\right)+P_b^{22}\Gamma^{1}{}_{22}+P_b^{33}\Gamma^{1}{}_{33}\nonumber\\
	&=\left(\rho_b+P_b\right)\Gamma^0{}_{10}+P_{b,1}+2\Gamma^{2}{}_{12}\left(P_b-P_{b_\perp}\right)
\end{align}which leads to the \emph{anisotropic} Tolman-Oppenheimer-Volkoff equation
\begin{equation}
	P_{b,1}+2\Gamma^{2}{}_{12}\left(P_b-P_{b_\perp}\right)=-\left(\rho_b+P_b\right)\Gamma^0{}_{10}\mdot\label{anTOVeq}
\end{equation}By specializing \eqref{anTOVeq} to our case, one finally finds
\begin{equation}
	{{v^2}}=\frac{Gm}{r-2Gm}\mdot
\end{equation}We then recover not only \emph{Keplerian  rotation} curves in the Newtonian limit ($v,Gm/r<<1$), but the \emph{full geodesic motion} of a test particle in the geometry generated by the surrounding dust swarm.
\paragraph{Coated case:}
In order to study the coated case we first proceed with the solution of $\varsigma_;{}^{\ti{S}}{}_{\ti{S}}=0$ in the static case, namely
\begin{align}
	\varsigma_;{}^{\ti{S}}{}_{\ti{S}}&=\varsigma{}_;{}^{1}{}_{1}=\varsigma_,{}^{1}{}_1+\varsigma_{,1}\Gamma^{01}{}_{0}\nonumber\\
	&=\mathrm{e}^{-\Phi-\Lambda}\frac{d}{dr}\left(\mathrm{e}^{\Phi-\Lambda}\frac{d}{dr}\varsigma\right)\eqimp 0
\end{align} so that
\begin{equation}
	\varsigma_{,1}=\pm\frac{\mathrm{e}^{-\Phi}}{\stackrel{\star}{R}}\label{solS}
\end{equation} for some scale factor $\stackrel{\star}{R}\;>0$. 
\paragraph{Komar mass:} The question now arises as to which sign to choose in expression \eqref{solS}. To this end, let us recall that we have a timelike Killing vector filed $\partial_t$. Thanks to the \emph{Ricci identity for Killing fields} and introducing the $1$-form $k_{\ddot{I}}$ canonically associated to our Killing vector field we can write
\begin{align}
	k_{\ddot{I};}{}^S{}_S&=-R_{\ddot{I}}{}^{\ddot{S}}k_{\ddot{S}}\mcomma
\end{align} see e.g. \cite{Wald} for reference. This enables one to define the notion of \emph{Komar mass}, which for stationary spacetimes is a conserved quantity associated to the time translations generated by $\partial_t$. In our case one can thus introduce the \emph{Komar mass density} expressed as
\begin{equation}
	8\pi G\;\rho_k=R_{\ddot{0}}{}^{\ddot{0}}k_{\ddot{0}}\mdot
\end{equation}
Thanks to the clear cut separation of the dark component of the Ricci tensor from the rest one can thus introduce the notion of \emph{dark Komar mass density} expressed as
 \begin{equation}
 8\pi G\;\stackrel{\varsigma}{\rho}_k=\stackrel{\varsigma}{R}{}_{\ddot{0}}{}^{\ddot{0}}k_{\ddot{0}}\mdot
 \end{equation}This can be rephrased more explicitly thanks to \eqref{coatRi} and noticing that $k_{\ddot{0}}=-\mathrm{e}^{\varsigma+\Phi}$
 \begin{equation}
 4\pi G\;\mathrm{e}^{-\varsigma}\stackrel{\varsigma}{\rho}_k=\pm\frac{\Gamma^{01}{}_0+2\Gamma^{21}{}_2}{\stackrel{\star}{R}}+\frac{\mathrm{e}^{-\Phi}}{\stackrel{\star}{R}{}^2}\mdot\label{Komar}
 \end{equation}
 If we require this quantity to be \emph{positive} over the whole chart without any condition on the function $m(r)$ we must then choose the positive solution for $\varsigma_{,1}$.

We can finally apply the conservation law of baryonic stress-energy \eqref{barcons} using its explicit form \eqref{coatdiv} and write the \emph{coated} anisotropic Tolman-Oppenheimer-Volkoff equation
\begin{equation}
P_{b,1}+2\Gamma^{2}{}_{12}\left(P_b-P_{b_\perp}\right)=-\left(\rho_b+P_b\right)\Gamma^0{}_{10}-2P_b\varsigma^{,1}+(-\rho_b+2P_{b\perp})\varsigma^{,1}\mcomma
\end{equation} which, applied to our fluid model, yields
\begin{equation}
	v^2=\left(\frac{Gm}{r-2Gm}+\frac{\mathrm{e}^{-\Phi}r^{3/2}}{\stackrel{\star}{R}\left(r-2Gm\right)^{1/2}}\right)\left(1+\frac{\mathrm{e}^{-\Phi}r^{3/2}}{\stackrel{\star}{R}\left(r-2Gm\right)^{1/2}}\right)^{-1}\mdot\label{rotcurv}
\end{equation}

 Let us now focus on the \emph{Newtonian r\'egime}. We define this as the limit in which the quantities $Gm/r,\Phi,r/\stackrel{\star}{R}<<1$ and all their combined higher integer powers are negligible with respect to the first. In this case, \eqref{rotcurv} can be expanded at first order as
\begin{equation}
	v^2=\frac{G\;m(r)}{r}+\frac{r}{\stackrel{\star}{R}}\mdot\label{Newtrot}
\end{equation} 
In this r\'egime we thus have a correction to the Keplerian behaviour of rotation curves. Interpreted from a Newtonian point of view, this is equivalent to a Newtonian dark matter density $\rho_{{}_{D_N}}$ taking the form
\begin{equation}
	\rho_{{}_{D_N}}=\frac{1}{2\pi G\stackrel{\star}{R}r}
\end{equation}
and corresponds to what one would get from the expression of the Komar dark matter density \eqref{Komar} in this limit. It is interesting to note that this essentially agrees with the central density profile found in computer simulations of cosmological halo formation in the \emph{cold dark matter paradigm}, this is known as the Navarro-Frenk-White profile see e.g. \cite{CDM} and references therein.

\paragraph{Baryonic Tully-Fisher relation:} Let us now analyse \eqref{Newtrot} in a region in which we suppose the rotation curve to be flat and the baryonic mass density to be negligible.
Let us for this purpose call $R_f$ the value of $r$ coordinate at which this region is situated, $v_f$ the value of said flat velocity and $m_b$ the total baryonic mass enclosed in the spherical region of radius $R_f$. The flatness condition  $v'=0$ then gives
\begin{equation}
	G\stackrel{\star}{R}m_b=R_f^2
\end{equation}and inserting this into \eqref{Newtrot} we finally end up with
\begin{equation}
	m_b=\frac{\stackrel{\star}{R}}{4G}v_f^4\mdot\label{bartully}
\end{equation}
We hence found that if a rotation curve is flat at the outskirts of a spherically symmetric astrophysical object coated by dark matter in the Newtonian r\'egime, then there is a universal relation between the total baryonic mass and the flat rotational velocity. This is a relation which is known to hold in a surprisingly wide range of astrophysical situations, we refer again to \cite{BarTully} for a full account of the observational peculiarities of this fact and references to further literature on the subject. In its wide survey of literature on the subject, \cite{NatureD} reports the value  for galaxies
\begin{equation}
	\stackrel{\star}{R}\;=\left(2.2\pm0.3\right)10^{27}m\mdot
\end{equation}This result was originally stated in terms of solar masses $M_\odot$. Here we used $2GM_\odot=2.9\; 10^3m$. To put things into perspective, this  roughly corresponds to twice the diameter of the \emph{observable universe}.

Notice that in our toy model we are only able to predict a universal \emph{proportionality} law, the proportionality constant, being an integration constant, has still to be determined empirically and depends in principle on the particular celestial object of application. The fact that a seemingly \emph{universal} constant is experimentally found might indicate (within the present scheme of thought) that the dark coating paradigm might have a range of application beyond spherical symmetry, thus predicting the \emph{same} coating constant $\stackrel{\star}{R}$ for the region of the universe we live in. The scale of $\stackrel{\star}{R}$ could be seen as a further tantalising evidence of this.

As a didactic idealised scenario, let us now determine the baryonic mass density which in the Newtonian r\'egime corresponds to a \emph{completely flat rotation curve.} It suffices to consider the flatness condition as a first order differential equation for the function $m(r)$, namely
\begin{equation}
	Gm'(r)=\frac{G\;m(r)}{r}-\frac{r}{\stackrel{\star}{R}}
\end{equation} which is easily solved by
\begin{equation}
	4\pi G\;\rho_b=\frac{v^2_f}{r^2}-\frac{2}{r\stackrel{\star}{R}}
\end{equation}for which \eqref{bartully} holds for the total mass enclosed in the ball of positive mass density. In this example we see how the centre of our spherical object is dominated by baryonic matter, while the outskirts are dominated by dark matter.  

\subsubsection{Schwarzschildean limit}
If in our toy model we take the $\rho_{{}_0}\longrightarrow0$ limit for the swarm of dust, while keeping the total enclosed mass $m(r)=m_b$ constant we clearly obtain the dark coating of the exterior Schwarzschild spacetime. This is to say, the Schwarzschild metric gets conformally rescaled by a conformal factor $\mathrm{e}^{2\varsigma}$, where
\begin{equation}
	{\varsigma}=\frac{r}{\stackrel{\star}{R}}+\frac{2 G m_b}{\stackrel{\star}{R}}ln\left(\frac{r}{2Gm_b}-1\right)
\end{equation}which is nothing but the so called \emph{tortoise coordinate}. It is then clear from \eqref{coatEin} that the conformal factor $\mathrm{e}^{2\varsigma}$ in general generates a curvature singularity at the (now shattered) Schwarzschild horizon. This problem is only apparent though. We must not forget that the \emph{compatibility condition} on $J$ still has to be satisfied. While this was no issue in the Newtonian r\'egime, it is one if we insist in approaching the Schwarzschild radius. To examine this more closely, let us explicitly write $\rho_D$, $J_D$ and $P_D$ in this context. From \eqref{coatEin} we deduce
\begin{align}
\stackrel{\varsigma}{\C{G}}{}^0{}_0&={\varsigma_{,1}}\left(-2\Gamma^{01}{}_0+4\Gamma^{21}{}_2+\varsigma_{,1}\right)\nonumber\\
\stackrel{\varsigma}{\C{G}}{}^0{}_1&=0\nonumber\\
\stackrel{\varsigma}{\C{G}}{}^1{}_1&=-2\;\varsigma^{,1}{}_1+{\varsigma_{,1}}\left(4\;\Gamma^{21}{}_2+3\;\varsigma_{,1}\right)\mdot
\end{align} and using 
\begin{align}
\Gamma^{01}{}_0&=\frac{Gm_b}{r^2\sqrt{1-\frac{2Gm_b}{r}}}\nonumber\\
\Gamma^{21}{}_2&=\frac{\sqrt{1-\frac{2Gm_b}{r}}}{r}\nonumber\\
\varsigma_{,1}&=\frac{1}{\stackrel{\star}{R}\sqrt{1-\frac{2Gm_b}{r}}}\mdot
\end{align}
we obtain
\begin{align}
-8\pi G\rho_D&=\mathrm{e}^{-\frac{2r}{\stackrel{\star}{R}}}\left(\frac{r}{2Gm_b}-1\right)^{-\frac{4 G m_b}{\stackrel{\star}{R}}}\left(-\frac{2Gm_b}{\stackrel{\star}{R}r^2\left(1-\frac{2Gm_b}{r}\right)}+\frac{4}{r\stackrel{\star}{R}}+\frac{1}{\stackrel{\star}{R}{}^2\left(1-\frac{2Gm_b}{r}\right)}\right)\nonumber\\
8\pi G J_D&=0\nonumber\\
8\pi G P_D&=\mathrm{e}^{-\frac{2r}{\stackrel{\star}{R}}}\left(\frac{r}{2Gm_b}-1\right)^{-\frac{4 Gm_b}{\stackrel{\star}{R}}}\left(\frac{2Gm_b}{\stackrel{\star}{R}r^2\left(1-\frac{2Gm_b}{r}\right)}+\frac{4}{r\stackrel{\star}{R}}+\frac{3}{\stackrel{\star}{R}{}^2\left(1-\frac{2Gm_b}{r}\right)}\right)\mdot
\end{align}The scalar curvature gains hence a singularity at $r=2Gm_b$. However the compatibility condition on $J_D$ demands that $\rho_D<0$ given that $P_D>0$ on the whole $r>2Gm_b$ region. Let us then rephrase the $\rho_D<0$ condition in terms of the dimensionless quantities $\mu\eqdef G m_b/\stackrel{\star}{R}$ and $x=r/\stackrel{\star}{R}$. One finds
\begin{equation}
	x^2+4x-10\mu>0\mdot
\end{equation} This means that our solution is not acceptable within a spherical region of coordinate radius
\begin{equation}
	\stackrel{\star}{r}{}=2\stackrel{\star}{R}\left(\sqrt{1+\frac{5\mu}{2}}-1\right)\mdot
\end{equation}For a typical ultra compact object of astrophysical relevance we have $\mu<<1$, so that for all practical purposes one has
\begin{equation}
	\stackrel{\star}{r}\;=\frac{5Gm_b}{2}\mcomma
\end{equation}just between the photon sphere and the original event horizon. We then have the significant result that the singular part of the solution is \emph{removed} by the compatibility condition on $J$. Physically this means that our dark coating acquires a region of instability around the horizon where no \emph{static} solution exists, just like what happens inside of the photon sphere for timelike dust particles. Following this analogy with the photon sphere we could hence call this the \emph{torsion sphere}. In FIG. $1.$ the latter is put to scale with the photon sphere and the original Schwarzschild horizon within.
\begin{center}
	\begin{figure}
		\includegraphics[width=0.8\textwidth]{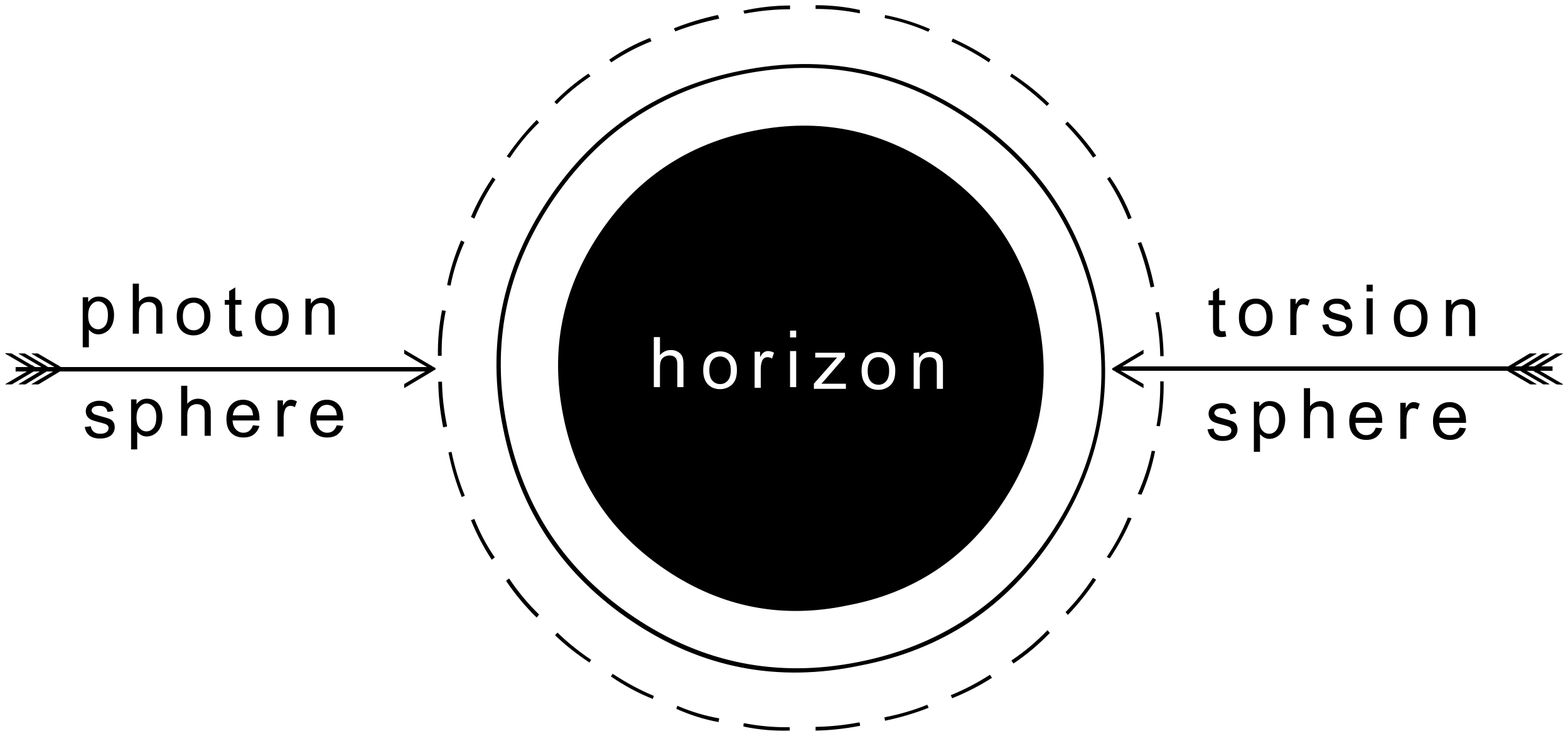}
		\caption{The torsion sphere put to scale with the photon sphere and event horizon}
	\end{figure}
\end{center}

Let us now turn to the rotation curves obtained in this limit setting. It suffices to specialise \eqref{rotcurv} to the case where $m(r)=m_b$. For later convenience, let us use the dimensionless quantities $\mu$ and $x$ as above. One has
\begin{equation}
	v^2=\frac{\mu+x^2}{x^2+x-2\mu}\mdot
\end{equation} It is then instructive to visualise this for a point-like analogue of the Milky Way, for which the greatest part of the baryonic matter is confined in a very compact region around the origin. In what follows we are only concerned by the qualitative and quantitative correspondence of out toy model with reality only as far as \emph{orders of magnitude} are concerned due to the idealised nature of our description. We refer to  \cite{Galac} for more details regarding astrophysical data. For our purposes it is sufficient to note that the ordinary baryonic content of our galaxy is estimated to be around $10^{11} M_\odot$ (compared to $10^{12}M_\odot$ obtained including dark matter-related phenomena), which corresponds to $\mu=6\cdot10^{-14}$. The rim of the galactic disc corresponds to a distance from the centre of the order of $6\cdot10^{20}m$, i.e. $x_r=3\cdot10^{-7}$. It is already evident that $x_r$ is of order $\sqrt{\mu}$ and that we thus are in the flat rotation curve region. In FIG. $2.$ we drew our rotation curve against its Keplerian relative (dashed line), highlighting the position that our sun would have. We thus find the orbital velocity in the outer region to be $8\cdot10^{-4}$, in S.I. units this is equivalent to $2\cdot10^5ms^{-1}$, exactly of the same order as the typical stellar velocity in the disc region of our galaxy. The picture turns then out to be consistent, with no need for exotic forms of matter.
\begin{center}
	\begin{figure}
		\includegraphics[width=0.8\textwidth]{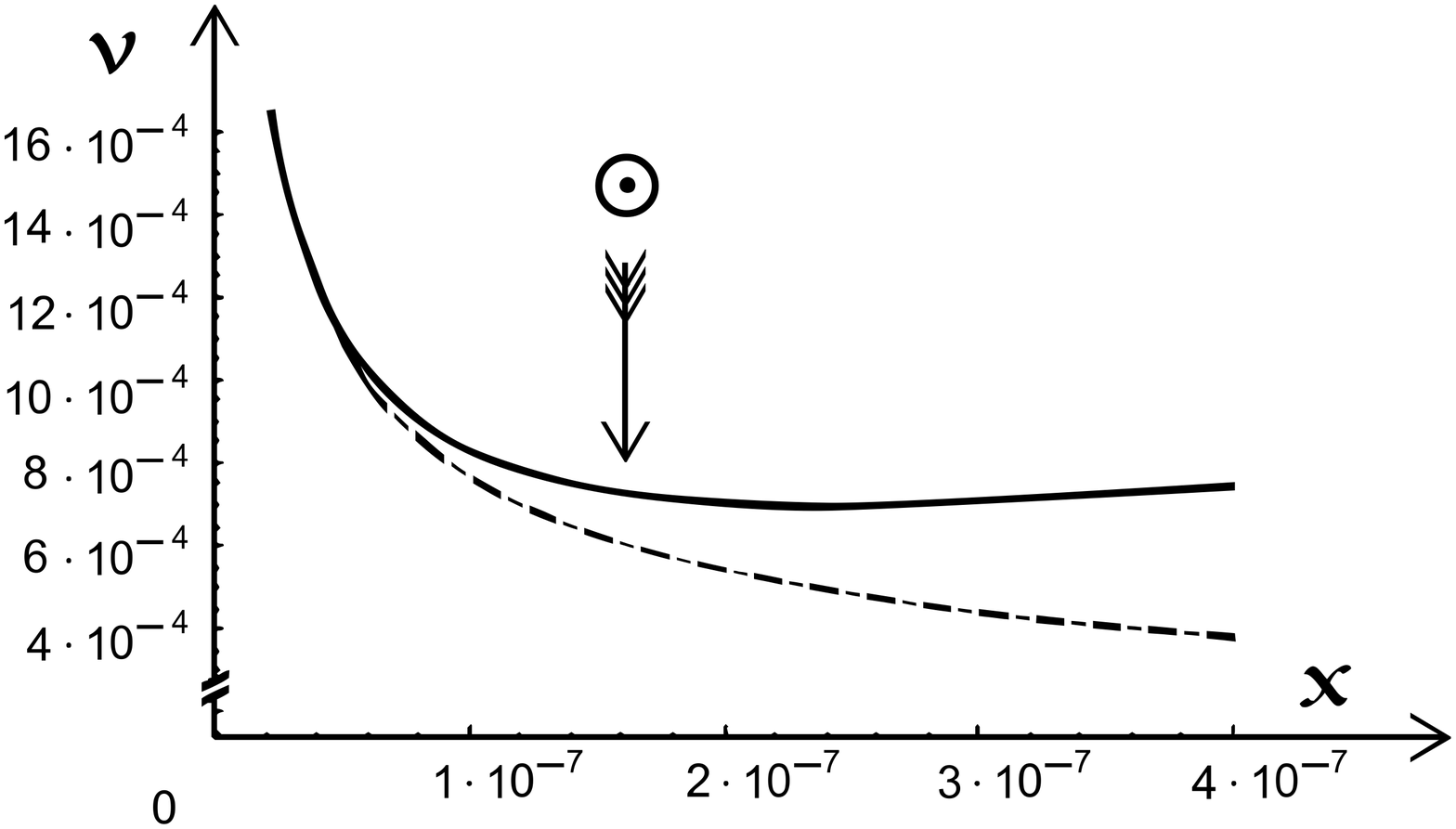}\label{Milk}
		\caption{Coated and uncoated rotation curves for a pointlike analogue Milky Way}
	\end{figure}
\end{center}
\subsection{Dynamical setting}
Let us now turn to a dynamical setting, i.e. devoid of timelike Killing fields. A general solution to \eqref{genwave} can be formally written in terms of local lightcone coordinates $(U,V)$ for the two dimensional Lorentzian metric $\f{\ti{g}_{0}}$ as follows:
\begin{equation}
	\varsigma=f(U)+g(V)
\end{equation} provided that the compatibility condition on $J$ be satisfied. In a Schwarzschildean setting $U$ and $V$ could for example be lightcone-adapted Kruskal-Szekeres coordinates. In such a case, all what has been previously said is recovered by choosing $f$ and $g$ to be equal and \emph{linear}. 
\subsubsection{Dark radiation}
In general, the $f$-component can be seen as an \emph{ingoing} spherically symmetric conformal wave, while the $g$-component as an \emph{outgoing} spherically symmetric conformal wave.
If one constructs a dark coating of the solution corresponding to a spherically symmetric pulsating star surrounded by vacuum one expects the boundary with the vacuum region to provide boundary conditions compatible with \emph{outgoing waves}. However, the problem of the details of e.g. stellar pulsations and that of the appropriate boundary conditions go beyond the scope of the present work. A useful exercise in that direction is to compute from \eqref{coatEin} the stress-energy tensor $P_g$ of a $g$ wave coating of the Schwarzschild exterior solution. From an uncoated perspective it is easy to see that such wave carries the following stress-energy tensor components
\begin{align}
	\rho_g{}&=\frac{-1}{4\pi G}
	\left(
		\frac{
				g{''}-g{'}{}^2-g{'}\frac{Gm_b}{r^2}
			}{1-\frac{2Gm_b}{r}}+\frac{2g{'}}{r}
	\right)\nonumber\\
		J_g&=\frac{1}{4\pi G}
	\left(
	\frac{
		-g{''}+g{'}{}^2+g{'}\frac{Gm_b}{r^2}
	}{1-\frac{2Gm_b}{r}}
	\right)\nonumber\\
	P_g&=\frac{1}{4\pi G}
	\left(
	\frac{
		-g{''}+g{'}{}^2+g{'}\frac{Gm_b}{r^2}
	}{1-\frac{2Gm_b}{r}}+\frac{2g{'}}{r}
	\right)\nonumber\\
		P_{g\perp}&=\frac{1}{4\pi G}
			\frac{g{'}}{r}
	\mcomma
\end{align}for which it is easy to see that the compatibility condition on $J$ always holds true. Notice how this stress-energy tensor has a \emph{traceless} far field limit which can be interpreted as the stress-energy tensor of a swarm of lightlike particles. This analogy becomes more evident if we consider a periodic $g$ with mean values over a period $T$ taking the values ${<g{'}>_T}={<g{''}>_T}=0$ and $<g{'}{}^2>_T=\gamma$ (e.g. a sine wave). Let us call $k^I$ the normalised future directed null vector pointing in the radial direction ($k^0=k^1=1$). Thanks to the time translation invariance of the uncoated base, it is meaningful to compute the period-averaged stress-energy tensor $<P_g{}^I{}_J>$ at every event. We find
\begin{equation}
	<P_g{}^I{}_J>_T=\frac{\gamma}{4\pi G}\frac{k^Ik_J}{\left(1-\frac{2Gm_b}{r}\right)}\mdot
\end{equation} This has thus the exact form of the stress energy tensor of a swarm of lightlike particles, whose associated wavenumber gets redshifted by the correct factor $\sqrt{1-\frac{2Gm_b}{r}}$.

This form of \emph{dark radiation} would be a promising observational signal and test of the present theory given its stark contrast with the Einsteinian setting, where no spherically symmetric gravitational radiation is present. 
\subsubsection{Hubble expansion}

As a final remark, it is interesting to note that a \emph{linear Hubble expansion} with Hubble constant $H$ can always be implemented by a dark coating by choosing linear functions $g$ and $f$ so that $f'=-g'=H$. If we take again the Schwarzschild spacetime and take an uncoated point of view one finds
\begin{align}
\rho_H{}&=\frac{1}{8\pi G}
\frac{
	3H^2
}{\left(1-\frac{2Gm_b}{r}\right)}\nonumber\\
J_H&=\frac{1}{4\pi G}
\frac{
	H\frac{Gm_b}{r^2}
}{\sqrt{1-\frac{2Gm_b}{r}}}\nonumber\\
P_H&=\frac{-1}{8\pi G}
\frac{
	H^2
}{\left(1-\frac{2Gm_b}{r}\right)}\nonumber\\
P_{H\perp}&=\frac{-1}{8\pi G}
\frac{
	H^2
}{\left(1-\frac{2Gm_b}{r}\right)}\mdot
\end{align}This satisfies the compatibility condition on $J$. Notice that it is an \emph{isotropic} stress-energy tensor, with equation of state $P=-\rho/3$ and \emph{negative} pressure. This is just the critical value of the perfect fluid constant $w=-1/3$ which underlies many cosmological discussions and makes room for constant expansion. Such possibilities are hence expected to be very promising on the cosmological arena and provide natural toy models of point-like perturbation of the FLRW cosmological paradigm \cite{Wein}.

\section{Conclusions and open questions}
The present theory seems to offer a promising framework for the interpretation of dark matter related phenomena in terms of geometry. In particular, the concept of conformal dark coating, pointed out by the symmetries of the equations of motion, seems of wide applicability, from galactic dynamics to cosmological scales. Further work has to be done in order to determine the robustness of this paradigm outside the spherically symmetric setting. Axially symmetric models will hopefully shed more light on galactic dynamics, thus enabling to produce actual fits of real rotation curves. The question of the non uniqueness of solutions, especially in the non static setting, remains wide open though. The question of possible realistic physical sources of dark radiation has also to be addressed through a thorough implementation of e.g. a stellar model involving pulsations able to generate such signals.
A further development of paramount importance will also be the application of the present ideas to cosmological dynamics, given that the theory so naturally includes isotropic Hubble expansion within its possibilities.

\bibliography{Bibliography}

\end{document}